\documentclass[english,twocolumn]{revtex4}
\usepackage[T1]{fontenc}
\usepackage[latin9]{inputenc}
\usepackage{amsmath}
\usepackage{color}
\usepackage{graphicx}

\makeatletter
\usepackage{babel}
\makeatother

\begin{document}

\title{Bell inequalities for continuous-variable correlations}

\author{E. G. Cavalcanti}

\affiliation{ARC Centre of Excellence for Quantum-Atom Optics, The University
of Queensland, Brisbane, Australia}

\author{C. J. Foster}

\affiliation{ARC Centre of Excellence for Quantum-Atom Optics, The University
of Queensland, Brisbane, Australia}

\author{M. D. Reid}

\affiliation{ARC Centre of Excellence for Quantum-Atom Optics, The University
of Queensland, Brisbane, Australia}

\author{P. D. Drummond}

\affiliation{ARC Centre of Excellence for Quantum-Atom Optics, The University
of Queensland, Brisbane, Australia}

\begin{abstract}
We derive a new class of correlation Bell-type inequalities. The inequalities
are valid for any number of outcomes of two observables per each of
$n$ parties, including continuous and unbounded observables. We show
that there are no first-moment correlation Bell inequalities for that
scenario, but such inequalities can be found if one considers at least
second moments. The derivation stems from a sim\textcolor{black}{ple
variance inequality by setting local commutators to zero. We show
that above a constant detector efficiency threshold, the continuous
variable Bell violation can survive even in the macroscopic limit
of large} $n$.\textcolor{black}{ This method can be used to derive
other well-known Bell inequalities, shedding new light on the importance
of non-commutativity for violations of local realism.}
\end{abstract}
\maketitle
Einstein, Podolsky and Rosen (EPR), in their famous 1935 paper \citep{Einstein1935},
demonstrated the incompatibility between the premises of \emph{local
realism} and the completeness of quantum mechanics. The original EPR
pape\textcolor{black}{r used continuous position and momentum variables,
and relied on their commutation relations, via the corresponding uncertainty
principle. Bohm \citep{Bohm1951} introduced, in 1951, his version
of the EPR paradox with spin observables. This was the version that
was used by Bell \citep{Bell1964} to prove his famous theorem showing
that quantum mechanics predicts results which can rule out the whole
class of local hidden variable (LHV) theories. It is hard to overemphasize
the importance of this result, which has even been called {}``the
most profound discovery of science'' \citep{Stapp1977}. However,
the original Bell inequality, and all of its generalizations, are
directly applicable only to the case of discrete observables. The
main purpose of this letter is to close the circle and derive a class
of Bell-type inequalities applicable to continuous variables correlations,
together with multipartite generalizations.}

\textcolor{black}{We derive a class of inequalities for local realism
that directly} use correlations of measurements, with no restriction
to spin measurements or discrete binning. The new inequalities are
remarkably simple. They place no restriction o\textcolor{black}{n
the number of possible outcomes, and the contrast between the classical
and quantum bounds involves commutation relations in a central way.
They must be satisfied b}y any observations in an LHV theory, whether
having discrete, continuous or unbounded outcomes. We can immediately
rederive previously known Bell-type inequalities, obtaining at the
same time their quantum-mechanical bounds by considering the non-commutativity
of the observables involved. We also display quantum states that directly
violate the new inequalities for continuous, unbounded measurements,
\textcolor{black}{even in the macroscopic, large} $n$ limit \citep{Mermin1990,Drummond1983,Peres1999,Reid2001}.
We show that \textcolor{black}{the new Bell violations survive the
effects of finite} generation and detection efficiency. This is very
surprising, in view of the many examples in which decoherence rapidly
destroys macroscopic superpositions \citep{Zurek2003}.

\textcolor{black}{Apart from this intrinsic interest, these inequalities
are relevant to an important scientific problem. No experiment has
yet produced a Bell inequality violation without introducing either
locality or detection loopholes. One path towards this goal is to
use continuous-variables (CV) and efficient homodyne detection, which
allows much higher detection efficiency than is feasible with discrete
spin or photo-detection measurements. A number of loop-hole free proposals
exist in the literature, but they all use Bell \citep{Leonhardt1995,Gilchrist1998,Auberson2002,Wenger2003,Garcia-Patron2004}
or Hardy \citep{Yurke1999} inequalities with a dichotomic binning
of the results (which usually lead to small violations), or else a
parity or pseudo-spin approach\citep{Banaszek1999,ChenZB2002a,Son2006b}
which cannot be realized with efficient homodyne detection. Are there
Bell inequalities which can be derived} \textcolor{black}{\emph{without}}
\textcolor{black}{the assumption of finite number of outcomes and
therefore are directly applicable to CV - with no need to bin the
results?}

For $n$ parties, $m$ measurements per party and $o$ outcomes, it
is well-known that the set of correlations allowed by LHV theories
can be represented as a \emph{convex polytope}, a multi-dimensional
geometrical structure formed by all convex combinations (linear combinations
where the coefficients are probabilities, i.e., they are non-negative
and sum to one) of a finite number of vertices. The vertices of this
polytope are the classical pure states --- the states with well-defined
values for all variables \citep{Pitowsky1989,Peres1999,Gisin2007}.
The tight Bell inequalities are associated with the linear facets
of the polytope. It is a computationally hard problem to list all
Bell inequalities for given $(n,m,o)$, and full numerical characterizations
have been accomplished only for small values of those parameters.

\textcolor{black}{However,} no class of Bell inequalities has previously
been derived without \emph{any} reference to the number of outcomes
or to their bound. Any real experiment will always yield a finite
number of outcomes; but are there constraints imposed by LHV theories
that are independent of any \emph{particular} discretization, and
can be explicitly written even in the limit $o\rightarrow\infty$?
Our answer is yes; and the derivation is much more straightforward
than in the case of the usual Bell-type inequalities, which are restricted
to a particular set of outputs.

We will focus on the \emph{correlation functions of observables}
for $n$ sites or observers, each equipped with $m$ possible apparatus
settings to make their causally separated measurements. We consider
any real, complex or vector function $F\left(\mathbf{X},\mathbf{Y},\mathbf{Z},\ldots\right)$
of local observations $X_{i},Y_{i},Z_{i}$ at each site $i$, which
in an LHV theory are all functions of hidden variables $\lambda$.
In a real experiment the different terms in $F$ may not all be measurable
at once, because they may involve different choices of incompatible
observables. The assumption of \emph{locality} enters the reasoning
by requiring that the local choice of observable does not affect the
correlations between variables at different sites, and therefore that
the averages are taken over the same hidden variable ensemble $P\left(\lambda\right)$
for all terms. We introduce averages over the LHV ensemble (there's
no loss of generality in considering deterministic LHVs \citep{Fine1982}),\begin{equation}
\left\langle F\right\rangle =\int P\left(\lambda\right)F\left(\mathbf{X}\left(\lambda\right),\mathbf{Y}\left(\lambda\right),\mathbf{Z}\left(\lambda\right),\ldots\right)d\lambda\,\,.\label{eq:LHV average}\end{equation}

Our LHV inequality uses the simple result that any function of random
variables has a non-negative variance, \begin{equation}
\left|\left\langle F\right\rangle \right|^{2}\leq\langle\left|F\right|^{2}\rangle.\label{eq:variance}\end{equation}
 We can also give a bound $\langle\left|F\right|^{2}\rangle\leq\langle\left|F\right|^{2}\rangle_{sup}$,
where the subscript denotes the supremum (least upper bound), in which
products of incompatible observables are replaced by their maximum
achievable values. This is necessary since if we are not able to measure
both $X_{i}$ and $Y_{i}$ simultaneously, a \textcolor{black}{general
LHV model could predict any achievable correlation \citep{Seevinck2007}.}

The same variance inequality applies to the corresponding Hermitian
operator $\hat{F}$ in quantum mechanics. While the observables at
different sites commute --- they can be simultaneously measured ---
those at the same site do not, so operator ordering must be included.
This enables us to see how quantum theory can violate the variance
bound for an LHV.

As an example, we will apply this variance inequality to a well-known
case. Consider two dichotomic observables $X_{i},Y_{i}$ per site
$i$, the outcomes of which are $\pm1$. We define $F_{1}\equiv X_{1}$,
$F_{1}'\equiv Y_{1}$ , and then inductively construct \citep{Gisin1998}:

\begin{equation}
F_{n}\equiv\frac{1}{2}(F_{n-1}+F_{n-1}')X_{n}+\frac{1}{2}(F_{n-1}-F_{n-1}')Y_{n},\label{eq:Fn}\end{equation}
where $F_{n}'$ can be obtained from $F_{n}$ by the exchange $X_{i}\longleftrightarrow Y_{i}$.
In calculating $F_{n}^{2}$ we'll keep track of the local commutators
just to make the contrast with quantum mechanics clearer. For real
variables $X,\, Y$, the commutator is defined in the same way as
for the corresponding operators, i.e., $[X,Y]\equiv XY-YX$. The anti-commutator
is defined by $[X,Y]_{+}\equiv XY+YX$. Then \begin{multline}
F_{n}^{2}=\frac{1}{4}\left\{ (F_{n-1}^{2}+F_{n-1}'^{2})(X_{n}^{2}+Y_{n}^{2})\right.\\
+[F_{n-1},F_{n-1}']_{+}(X_{n}^{2}-Y_{n}^{2})+(F_{n-1}^{2}-F_{n-1}'^{2})[X_{n},Y_{n}]_{+}\\
\left.-[F_{n-1},F_{n-1}'][X_{n},Y_{n}]\right\} .\label{eq:Fn2}\end{multline}

Since $\hat{X}_{n}^{2}=\hat{Y}_{n}^{2}=1$, we can show that \textcolor{black}{$F_{n}^{2}=F_{n}'^{2}$
and \begin{equation}
F_{n}^{2}=F_{n-1}^{2}-\frac{1}{4}\left[F_{n-1},F_{n-1}'\right]\left[X_{n},Y_{n}\right].\label{eq:FN2 - 2}\end{equation}
}In a LHV theory, the term which involves commutators will be zero
since $[X(\lambda),Y(\lambda)]=X(\lambda)Y(\lambda)-Y(\lambda)X(\lambda)=0$.
H\textcolor{black}{ence by induction $F_{n}^{2}=F_{1}^{2}=1$ and
the variance inequality (\ref{eq:variance}) becomes:} $-1\leq\left\langle F_{n}\right\rangle \leq1.$
This is the Mermin-Ardehali-Belinskii-Klyshko (MABK) \citep{Mermin1990,Ardehali1992,Belinskii1993}
Bell inequality, which reduces to the well-known Bell-CHSH \citep{Clauser1969}
inequality for $n=2$. 

We can now calculate the quantum mechanical bound by writing the variance
inequality \eqref{eq:variance} and substituting the functions in
\eqref{eq:FN2 - 2} by their corresponding operators\begin{eqnarray}
\bigl\langle\hat{F}_{n}\bigr\rangle_{Q}^{2} & \leq & \bigl\langle\hat{F}_{n}^{2}\bigr\rangle_{Q}=\bigl\langle\hat{F}_{n-1}^{2}-\frac{1}{4}[\hat{F}_{n-1},\hat{F}_{n-1}'][\hat{X}_{n},\hat{Y}_{n}]\bigr\rangle_{Q}\nonumber \\
 & \leq & \bigl\Vert\hat{F}_{n-1}^{2}\bigr\Vert+\frac{1}{4}\bigl\Vert[\hat{F}_{n-1},\hat{F}_{n-1}']\bigr\Vert\bigl\Vert[\hat{X}_{n},\hat{Y}_{n}]\bigr\Vert,\label{eq:QM Bound 1}\end{eqnarray}
where the norm $\left\Vert A\right\Vert $ denotes the modulus of
the maximum value of $\langle\hat{A}\rangle_{Q}$ over all quantum
states. The norm of the second commutator has the bound $\Vert[\hat{X}_{n},\hat{Y}_{n}]\Vert\leq2$.
It's easy to show that $[\hat{F}_{n},\hat{F}_{n}']=\hat{F}_{n-1}^{2}[\hat{X}_{n},\hat{Y}_{n}]+[\hat{F}_{n-1},\hat{F}_{n-1}']$
and therefore $\Vert[\hat{F}_{n},\hat{F}_{n}']\Vert\leq2\Vert\hat{F}_{n-1}^{2}\Vert+\Vert[\hat{F}_{n-1},\hat{F}_{n-1}']\Vert$
. Solving the recursion relation by noting that $\Vert\hat{F_{1}^{2}}\Vert=\frac{1}{2}\Vert[\hat{X}_{1},\hat{Y}_{1}]\Vert=1$
we finally arrive at the bound $\langle\hat{F_{n}}\rangle_{Q}^{2}\leq2^{n-1}.$
This can be attained with the generalized GHZ states \citep{Gisin1998},
which therefore violate (\ref{eq:variance}).

Inspired by those results, we now demonstrate an LHV inequality that
is directly applicable to unbounded continuous variables, in particular
field quadrature operators. The choice of the function $F_{n}$ in
(\ref{eq:Fn}) is not optimal though, since the variance in general
involves incompatible operator products that have no upper bound.

\textcolor{black}{To overcome this problem, consider a complex function
$C_{n}$ of the local real observables $\{X_{k},\, Y_{k}\}$ defined
as:\begin{equation}
C_{n}=\tilde{X_{n}}+i\tilde{Y_{n}}=\prod_{k=1}^{n}(X_{k}+iY_{k})\,,\end{equation}
 so that the modulus square only involves compatible operator products,
i.e. $\left|C_{n}\right|^{2}=\prod_{k=1}^{n}(X_{k}^{2}+Y_{k}^{2})\,.$
Applying the variance inequality to both $\tilde{X_{n}}$ and $\tilde{Y_{n}}$,
we find that: }

\textcolor{black}{\begin{equation}
\langle\tilde{X}_{n}\rangle^{2}+\langle\tilde{Y_{n}}\rangle^{2}\leq\langle\prod_{k=1}^{n}(X_{k}^{2}+Y_{k}^{2})\rangle\label{eq:New Ineq final}\end{equation}
This is our main result. Given the assumption of local hidden variables,
this inequality must be satisfied for any set of observables $X_{k}$,
$Y_{k}$, regardless of their spectrum. }

The fact that we have neglected the commutators in deriving \eqref{eq:New Ineq final}
hints that quantum mechanics might predict a violation. We define
quadrature operators\begin{eqnarray}
\hat{X}_{k} & = & \hat{a}_{k}e^{-i\theta_{k}}+\hat{a}_{k}^{\dagger}e^{i\theta_{k}}\nonumber \\
\hat{Y}_{k} & = & \hat{a}_{k}e^{-i(\theta_{k}+s_{k}\pi/2)}+\hat{a}_{k}^{\dagger}e^{i(\theta_{k}+s_{k}\pi/2)},\label{eq:XkYk}\end{eqnarray}
where $\hat{a}_{k},\hat{a}_{k}^{\dagger}$ are the boson annihilation
and creation operators at site $k$ and $s_{k}\in\{-1,1\}$. 

We now define the \textcolor{black}{operator $\hat{Z}_{k}\equiv\hat{X}_{k}+i\hat{Y}_{k}$
and note that it follows that $\hat{C_{n}}=\prod_{k=1}^{n}\hat{Z}_{k}$.}
The definition of $\hat{Y}_{k}$ allows for the choice of the relative
phase with respect to $\hat{X}_{k}$ to be $\pm\pi/2$. Depending
on $s_{k}$, for each $k$ \textcolor{black}{either $\hat{Z}_{k}=2\hat{a}_{k}e^{-i\theta_{k}}$
or $\hat{Z}_{k}=2\hat{a}_{k}^{\dagger}e^{i\theta_{k}}$. Denoting
$\hat{A}_{k}(1)=\hat{a}_{k}$ and $\hat{A}_{k}(-1)=\hat{a}_{k}^{\dagger}$,
the term in the LHS of (\ref{eq:New Ineq final}) in quantum mechanics
is then $|\langle\prod_{k}\hat{Z}_{k}\rangle_{Q}|^{2}=|2^{n}e^{i\sum_{k}s_{k}\theta_{k}}\langle\prod_{k}\hat{A}_{k}(s_{k})\rangle_{Q}|^{2}$.
The RHS becomes $\langle\prod_{k=1}^{n}(4\hat{a}_{k}^{\dagger}\hat{a}_{k}+2)\rangle_{Q}$
regardless of the phase choices. To violate (\ref{eq:New Ineq final})
we must therefore find a state that satisfies}

\textcolor{black}{\begin{equation}
\Bigl|\Bigl\langle\prod_{k}\hat{A}_{k}(s_{k})\Bigr\rangle_{Q}\Bigr|^{2}>\Bigl\langle\prod_{k}\bigl(\hat{a}_{k}^{\dagger}\hat{a}_{k}+\frac{1}{2}\bigr)\Bigr\rangle_{Q},\label{eq:New Ineq a}\end{equation}
}which is surprisingly insensitive to relative phases between the
quadrature measurements at different sites.

This violation of a continuous variable Bell inequality can be realized
within quantum mechanics. Consider an even \textcolor{black}{number
of sites, choosing $s_{k}=1$ for the first half of them and $s_{k}=-1$
for the remaining. To maximize the LHS we need a superposition of
terms which are coupled by that product of annihilation/creation operators.
One choice is a state of type}

\textcolor{black}{\begin{equation}
\left|\Psi_{S}\right\rangle =c_{0}\left|0,\dots,0,1,\dots,1\right\rangle +c_{1}\left|1,\dots,1,0,\dots,0\right\rangle ,\label{eq:state}\end{equation}
where in the first term the first $n/2$ modes are occupied by zero
photons and the remaining by $1$; conversely for the second term.
With that choice of state the LHS of \eqref{eq:New Ineq a} becomes
$|c_{0}|²|c_{1}|²$, which is maximized by $|c_{0}|^{2}=|c_{1}|^{2}=\frac{1}{2}$.
The RHS is $(\frac{3}{2})^{\frac{n}{2}}(\frac{1}{2})^{\frac{n}{2}}$
independently of the amplitudes $c_{0},\, c_{1}$. Dividing the LHS
by the RHS, inequality (\ref{eq:New Ineq a}) becomes $\frac{1}{4}\left(\frac{4}{3}\right)^{\frac{n}{2}}\leq1$,
which is vio}lated for $n\geq10$, and the violation grows exponentially
with the number of sites. 

While setting up the homodyne detectors necessary for this observation
is challenging, the complexity of this task scales linearly with the
number of modes. A more stringent constraint is most likely in the
state preparation, but we can relate state (\ref{eq:state}) to a
class of states of great experimental interest. They can be achieved
from a generalized GHZ state of $n/2$ photons, $\frac{1}{\sqrt{2}}(|H\rangle^{\otimes\frac{n}{2}}+|V\rangle^{\otimes\frac{n}{2}})$
--- where $|H\rangle$ and $|V\rangle$ respectively represent single-particle
states of horizontal and vertical polarization --- by splitting each
mode with a polarizing beam splitter. Therefore violation of (\ref{eq:New Ineq final})
can be observed in the ideal case with a 5-qubit photon polarisation
GHZ state and homodyne detection. 

An interesting question is the effect of decoherence, both from state
preparation error \citep{Jang2006} and detector inefficiency. The
usual Bell-CHSH violations have an efficiency threshold \citep{Garg1987}
of $83\%$. This has not yet been achieved for single-photon counting.
Homodyne detection is remarkably efficient by comparison, with up
to $99\%$ efficiencies being reported. However, the effect of detector
efficiency is easily included by assuming that each detected photon
mode is preceded by a beamsplitter with intensity transmission $\eta<1$.
This changes both the LHS and RHS, so that the inequality becomes
\textcolor{black}{$\frac{4\eta^{2}}{2\eta+1}\leq4^{2/n}$, giving
a threshold efficiency requirement of $\eta>\eta_{min}$, where $\eta_{min}=(1+\sqrt{1+4^{1-2/n}})/4^{1-2/n}$. }

This \emph{reduces} at large $n$ to an asymptotic value of $\eta_{\infty}=0.80902$.
Unexpectedly, the Bell violation (which signifies a quantum superposition)
is less sensitive to detector inefficiency in the macroscopic, large
$n$ limit. The minimum detector efficiency $\eta_{n}$ at finite
$n$ is plotted in Fig. 1, together with the minimum state preparation
fidelity $\epsilon_{min}$ in the case of ideal detectors, where we
model the density matrix as $\hat{\rho}=\epsilon|\Psi_{S}\rangle\langle\Psi_{S}|+(1-\epsilon)\hat{I}$.

\begin{figure}
\includegraphics[scale=0.45]{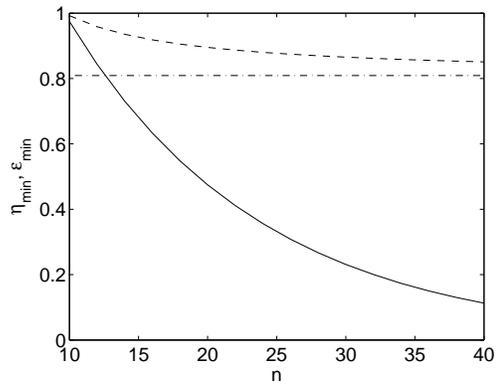}

\caption{Minimum state preparation fidelity $\epsilon_{min}$ for ideal detectors
(solid line), and minimum detection efficiency $\eta_{min}$ for ideal
state preparation (dashed line) required for violation of (\ref{eq:New Ineq final})
as a function of the number of modes. The asymptotic value of $\eta_{min}$
is indicated by the dash-dotted line.}

\end{figure}

We will finally prove that there are no LHV inequalities possible
if one considers only the first-moment correlations between continuous
variables in different sites. We will show this explicitly for the
simplest case and indicate how to generalize to arbitrary numbers
of parties and settings. Consider first $n=2$ parties, Alice and
Bob, each of which can choose between $m=2$ observables: $X_{a},Y_{a}$
for Alice and $X_{b},Y_{b}$ for Bob. Each measurement yields an outcome
in the real numbers. The first-moment correlation functions for each
of the $4$ possible configurations are just the averages $\langle X_{a}X_{b}\rangle$,
$\langle X_{a}Y_{b}\rangle$, $\langle Y_{a}X_{b}\rangle$, $\langle Y_{a}Y_{b}\rangle$.
Given those $4$ experimental outcomes, can we find a local hidden
variable model which reproduces them? 

\textcolor{black}{We construct an explicit example. Consider a hidden-variable
state $S$ where the hidden variables are the measured values $\mathbf{X},\,\mathbf{Y}$,
in an equal mixture of four classical pure} \textcolor{black}{states
$S_{k}=\left(X_{a},Y_{a},X_{b},Y_{b}\right)_{k}$ defined by}

\textcolor{black}{\begin{equation}
\begin{split}S_{1} & =2\,(1,0,\langle X_{a}X_{b}\rangle,0)\\
S_{2} & =2\,(1,0,0,\langle X_{a}Y_{b}\rangle)\\
S_{3} & =2\,(0,1,\langle Y_{a}X_{b}\rangle,0)\\
S_{4} & =2\,(0,1,0,\langle Y_{a}Y_{b}\rangle).\end{split}
\label{eq:stateSdef}\end{equation}
}

\textcolor{black}{Each of the states $S_{k}$ assigns a nonzero value
to only one of the $4$ correlation functions. Since the probability
of each of the states in the equal mixture is $1/4$, we have for
example $\langle X_{a}X_{b}\rangle_{S}=\frac{1}{4}\sum_{i}\langle X_{a}X_{b}\rangle_{S_{i}}=\langle X_{a}X_{b}\rangle$.}

\textcolor{black}{Satisfying the two-site correlations using the state
$S$ defined by \eqref{eq:stateSdef} leaves us with uncontrolled
values for the single-site correlations, for instance $\langle X_{b}\rangle_{S}=\frac{1}{2}(\langle X_{a}X_{b}\rangle+\langle Y_{a}X_{b}\rangle)$.
One might object to the fact that this is not equal to $\langle X_{b}\rangle$
in general. However, we may correct these lower order correlations
by adding four more states ($S_{5}$ to $S_{8}$) and changing the
prefactors multiplying $S_{1}$ to $S_{4}$ to compensate for their
reduced weight in the equal mixture. Crucially, adding these extra
states to $S$ in this manner does not modify the values of correlations
such as $\langle X_{a}X_{b}\rangle$. As an example, we exhibit the
state $S_{5}=8\,(0,0,\langle X_{b}\rangle-(\langle X_{a}X_{b}\rangle+\langle Y_{a}X_{b}\rangle)/\sqrt{8},0)$,
which corrects the single expectation value $\langle X_{b}\rangle_{S}$
to $\langle X_{b}\rangle$. }

\textcolor{black}{The proof generalizes easily to arbitrary $n$ and
$m$. In that case, there are $m^{n}$ possible combinations of measurements
which yield $n$-site correlations. Denoting the $j^{\mathrm{th}}$
observable at site $i$ by $X_{i}^{j}$, each combination is specified
by a sequence of indices $(j_{1},j_{2},\dotsc,j_{n})$. For each combination
of measurements, we define a hidden variable state which assigns nonzero
values only to the variables which appear in the associated correlation
function $\langle\prod_{i=1}^{n}X_{i}^{j_{i}}\rangle$. In analogy
to the example above, we can always choose the values of the hidden
variables associated to $X_{i}^{j_{i}}$ such that their product is
equal to $m^{n}\langle\prod_{i=1}^{n}X_{i}^{j_{i}}\rangle$. Since
all other $m^{n}-1$ states defined in this way will give a value
of zero to this particular correlation function, and given that the
probability associated with each of those states is $1/m^{n}$, we
reproduce all correlations as desired. As indicated in the example,
additional first moment correlations involving} \textcolor{black}{\emph{less}}
\textcolor{black}{than $n$ sites can be included in the LHV model
by adding additional states to $S$ in a way which doesn't affect
the $n$-site correlations. Thus, any possible observation of first
moment correlations may be explained using a LHV model, and hence
these correlations alone cannot violate any Bell inequality. In other
words, the minimum requirement for a correlation Bell inequality with
continuous, unbounded variables, is to use not just the first but
also the second moments at each site.}

In conclusion, we have derived a new class of Bell-type inequalities
valid for continuous and unbounded experimental outcomes. We have
shown that the same procedure allows one to derive the MABK class
of Bell inequalities and their corresponding quantum bounds. That
derivation makes it explicit that non-zero commutators --- associated
with the incompatibility of the local observables --- are the essential
ingredient responsible for the discrepancy between quantum mechanics
and local hidden variable theories. The new Bell-type inequality derived
here can be directly applied to continuous variables without the need
for a specific binning of the measurement outcomes. Surprisingly,
quantum mechanics predicts exponentially increasing violations of
the inequality for macroscopically large numbers of sites, even including
realistic decoherence effects like inefficient state preparation,
and a detector loss at \emph{every} site.

We thank Y.C. Liang and B. Lanyon for interesting and helpful discussions
and acknowledge the ARC Centre of Excellence program for funding this
research.

\end{document}